\documentclass[aps,amsmath,amssymb]{revtex4}

\usepackage{graphicx}
\usepackage{color}
\usepackage{dcolumn} 
\usepackage{bm}
\usepackage{multirow}
\usepackage{array}
\usepackage{bigdelim}

\begin{document}

\title{Dislocation mutual interactions mediated by mobile impurities
and the conditions for plastic instabilities}

\author{Fabio Leoni}
\affiliation{School of Mechanical Engineering, Tel-Aviv University,
  Tel-Aviv 69978, Israel} 
\author{Stefano Zapperi}
\affiliation{CNR-IENI, Via R. Cozzi 53, 20125 Milano, Italy\\
ISI Foundation, Via Alassio 11C, 10126 Torino, Italy}

\begin{abstract}
Matallic alloys, such as Al or Cu, or mild steel, display plastic
instabilities in a well defined range of temperatures and deformation rates,
a  phenomenon known as the Portevin-Le Chatelelier (PLC) effect.
The stick-slip behavior, or serration, typical of
this effect is due to the discontinuous motion of dislocations
as they interact with solute atoms.
Here we study a simple model of interacting 
dislocations and show how the classical Einstein fluctuation-dissipation
relation can be used to define the temperature in a range of model
parameters and to construct a phase diagram of serration that can be
compared to experimental results. Furthermore, performing analytical
calculations and numerically integrating the equations of motion, we 
clarify the crucial role played by dislocation mutual interactions in
serration.
\end{abstract}
 
\maketitle

\section{Introduction}
Dislocation dynamics is a complex intermittent phenomenon involving
the collective motion of many dislocations interacting between each
other as well with obstacles eventually present in the material, like
solute atoms or quenched dislocations from other glide planes
\cite{zaiser2006,weiss1997,miguel2001}. The long range stress
produced by dislocations may lead to jamming and avalanche-like
phenomena even in the absence of obstacles \cite{miguel2002}.
The presence of obstacles changes the local
properties of the host material, resulting in a pinning force on
nearby dislocations \cite{butt1993,neuhauser1993}. 
Usually, this source of disorder for dislocations is taken to be
quenched, so that its properties do not change within the
relevant timescales of the system \cite{leoni2009}. 
However, under specific conditions, the mobility of solute atoms in
metallic alloys \cite{cahn1962,blavette1999}, or oxygen vacancies in
superconductors \cite{chudnovsky1998}, plays an
important role in the dynamics of these systems.  

Here we are interested in studying the dynamics of interacting
dislocations mediated by mobile impurities. The interplay
between dislocations mutual interactions and pinning by mobile impurities is believed to be at the
origin of plastic instabilities observed in metallic alloys under suitable loading conditions 
and temperature. One of the best studied forms of instability propagation is the
Portevin-Le Chatelier (PLC) effect \cite{portevin1923}.
When a specimen of a dilute alloy (such as Al or Cu alloy, or mild
steel) is strained in uniaxial loading, the mechanical response is
often discontinuous. In constant applied strain rate tests, the stress
versus strain (or time, which is proportional to strain) curves
exhibit a succession of stress drops and reloading sequences
(serration).

From a dynamical point of view, the jerky or stick-slip behavior of
stress is related to the discontinuous motion of dislocations, namely,
the pinning (stick) and unpinning (slip) of dislocations. The
 classical explanation of the PLC effect is via the dynamic
strain aging (DSA) concept \cite{cottrell1949}. It is based on the
interplay between the diffusivity of solute atoms and dislocations
that can be arrested temporarily at obstacles during their waiting
time. Thus, the longer the dislocations are arrested, the larger will
be the stress required to unpin them. As a result, when the
contribution from aging is large enough, the critical stress to move a
dislocation increases with increasing waiting time or decreasing
imposed strain rate. When these dislocations are unpinned, they move at
large speed until they are arrested again. At high strain rates (or
low temperatures), the time available for solute atoms to diffuse towards
the dislocations in order to age them decreases and hence the stress required
to unpin them decreases. Thus, in a range of strain rates and
temperatures where these two time scales are of the same order of
magnitude, the PLC instability manifests. The competition between the
slow rate of aging and sudden unpinning of the dislocations,
translates at the macroscopic level, into a negative strain rate
sensitivity (SRS) of the flow stress as a function of strain
rate. This basic instability mechanism, used in most
phenomenological models for the PLC effect \cite{ananthakrishna2007},
is based on the behavior of individual dislocation and thus does not
explain how dislocation 
motion can synchronize to yield macroscopic strain bursts.

Current theoretical approaches to models the PLC effect are based on a
mesoscopic descriptive level (where coarse grained dislocation
density is considered) in which phenomenological parameters are
needed to construct the relative dynamical equations \cite{ananthakrishna2007}. 
Modeling plastic deformation phenomena taking into account inhomogeneity at the
dislocation level offers fundamental advantages compared to continuum mechanics
approaches.
The discrete dislocation dynamical (DDD) approaches allow, for
example, to account for the intrinsic length scales, such as the grain
size, the mesh length of a dislocation network or the cross-slip
height, which is necessary to understand the formation of spatial
dislocation structures, as persistent slip bands \cite{neuhauser1983},
and plastic instabilities, as L\"{u}ders band and the PLC effect
\cite{ananthakrishna2007}.  
The problem of spatial and temporal coupling in heterogeneously
deforming materials, and the associated length and time scales to be
included in constitutive laws, is a central issue in current attempts
to bridge a gap between dislocation based constitutive models and
continuum mechanics. 

The general three-dimensional dynamical problem of dislocation lines
interacting among each other is a complex problem due to the necessity
to consider flexible lines conserving their connectivity and line
length and take care of line interactions
\cite{mohles2001,monnet2006,csikor2007,devincre2008,senger2008,liu2011,reddy2013}.    
In several instances, however, dislocations are arranged into regular
structures that are amenable to analytical treatment and more
efficient simulation approach.
Here we analyze the dynamics of the effective one-dimensional
dislocation array called pileup interacting with mobile impurities.
A slip band can be envisaged as a queue of dislocations, a pileup,
pushed through a series of obstacles (solute atoms or immobile
dislocations from other glide planes). In our case the obstacles
perform a diffusive motion due to thermal effects, and interact with
dislocations. 
This system can be viewed as a coupled one-dimensional channels of particles 
\cite{bairnsfather2011}, in which particles in one channel (dislocations) 
are driven by an external force and experience a drag from the undriven particles 
(impurities) in the other channel.   
In the following we discuss before the single dislocation in a cloud
of mobile impurities problem, analyzed in Ref.\cite{laurson2008}, and
then we propose a generalization of the equations in the case of many
interacting dislocations in a landscape of mobile impurities. 

\section{Single dislocation interacting with mobile impurities}

Recently the dynamics of a particle interacting with diffusing
impurities in one dimension has been investigated by Laurson and Alava
\cite{laurson2008}. Despite the simplicity of the model, that makes it
analytically tractable, it exhibits a rich dynamics. Here we describe
this model as an introduction to the following section in which we
will generalize the relative equations to the case of many interacting
dislocations.  

In the full formulation of the model discussed in Ref.\cite{laurson2008},
a particle in a cloud composed by a fixed number of $N_p$ impurities
driven by an external force $F$ is considered.   
The force is given by $F=k(Vt-x)$, where $V$ is the driving velocity and $k$
is a spring constant characterizing the response of the driving
mechanism.
The region of the parameter space is restricted to that in which the
impurities have a vanishingly small probability to escape from the
vicinity of the particle within the timescale of the simulation. Thus,
the particle is dragging an impurity cloud with a fixed number of
impurity particles without escaping from it.
The equations of motion are
\begin{equation}\label{equ:laurson}
\begin{array}{lll}
\mu\dfrac{\partial x}{\partial t} & = &
\sum_{i=1}^{N_p}f(x-x_{s,i})+F ,\\
&&\\
\dfrac{\partial x_{s,i}}{\partial t} & = & -f(x-x_{s,i})+\eta_i ,
\end{array}
\end{equation} 
where $x$ and $x_s$ are the position of the particle and the impurity
particles, respectively.
$f(z)$ is the interaction force between the particle and the impurity particle,
$\mu$ defines the relative mobility of the impurity and the particle and
$\eta_i$ are Gaussian white noise with standard deviation $\delta\eta$
and zero mean.
The only condition imposed on the expression of the force $f(z)$ if
$\partial_zf(z)|_{z=0}=-f_0$. 
Here we are interested in particular on the behavior of the external force $F$,
that in experiments represents the shear stress acting on dislocations.
For $z=x-x_s$ close to zero, the following expression for the stochastic
process $\partial_t F$ is derived in Ref.~\cite{laurson2008}
\begin{equation}\label{equ:F}
\partial_t^2F=-k\partial_t^2x=-\left[\dfrac{k}{\mu}+\dfrac{f_0}{\mu}
(N_p+\mu)\right]\partial_tF+\dfrac{kf_0}{\mu}\sum_{i=1}^{N_p}\eta_i+\dfrac{kf_0}{\mu}\left[V(N_p+\mu)-F\right] . 
\end{equation} 
Now, assuming that in the stationary state the last term in the r.h.s.
of Eq.\ref{equ:F} has zero mean ($\langle V(N_p+\mu)-F\rangle=0$) and
that fluctuations are small compared to those of the white noise term ($\delta
F\ll\sqrt{N_p}\delta\eta$), the equation \ref{equ:F} reduces to the
following Ornstein-Uhlenbeck process for $\partial_tF$
\begin{equation}\label{equ:F2}
\partial_t^2F=-\left[\frac{k}{\mu}+\dfrac{f_0}{\mu}(N_p+\mu)\right
]\partial_tF+\dfrac{kf_0}{\mu}\sum_{i=1}^{N_p}\eta_i .
\end{equation} 
The condition of small fluctuations $\delta F\ll\sqrt{N_p}\delta\eta$  is
fulfilled for most of the relevant parameter values condition; only for
$kf_0\gg1$ this is not the case.
From Eq.~\ref{equ:F2} is possible to see that, after an initial transient, the
system reaches the stationary state in which the external force $F$
fluctuates around a constant average value and these fluctuations are
uncorrelated in time \cite{laurson2008}. 
Therefore, the system composed by a single dislocation in a cloud of mobile
impurities does not display a serration type behavior. 

The Eq.~\ref{equ:F} can be solved exactly, without imposing conditions
on fluctuations. Indeed is possible to rewrite it as a two dimensional
(2d) Ornstein-Uhlenbeck process. 
Introducing the new variable $F^*=F-V(N_p+\mu)$ and considering $F^*$ and
$\dot{F^*}$ as the components of a 2d vector, the Eq.~\ref{equ:F} can be written
as a 2d Ornstein-Uhlenbeck process \cite{risken1984} for the vector
variable $(F^*,\dot{F^*})$
\begin{equation}
\dfrac{d}{dt}\left(
\begin{array}{l} 
F^*\\ \dot{F^*}
\end{array}\right) = -\left(
\begin{array}{lr} 
0 & -1\\ \omega_0^2 & \gamma
\end{array}\right)\left(\begin{array}{l} 
F^*\\ \dot{F^*}
\end{array}\right)+\left(\begin{array}{l} 
0\\ \Gamma(t)
\end{array}\right)
\end{equation} 
where 
\begin{equation}
\begin{array}{rll}
\gamma & = & \dfrac{1}{\mu}[k+f_0(N_p+\mu)],\\
&&\\
\omega_0^2 & = & \dfrac{kf_0}{\mu},\\
&&\\
\Gamma(t) & = & \dfrac{kf_0}{\mu}\sum_{i=1}^{N_p}\eta_i.
\end{array}
\end{equation}
The solution for the average $\langle F^*\rangle$ is
\begin{equation}\label{G}
\langle F^*\rangle = e^{-{\bm \gamma}t}\left|_{11}\right.F^*(0)+e^{-{\bm
    \gamma}t}\left|_{12}\right.\dot{F^*}(0), 
\end{equation} 
where
\begin{equation}
\begin{array}{l}
F^*(0)=F(0)-V(N_p+\mu)=-V(N_p+\mu),\\
\\
\dot{F^*}(0)=\dot{F}(0)=kV,
\end{array}
\end{equation} 
and the matrix ${\bm\gamma}$ is 
\begin{equation}
{\bm\gamma}=\left(
\begin{array}{lr}
0 & -1\\
\omega_0^2 & \gamma
\end{array}\right) 
\end{equation} 
Diagonalizing the exponential matrix $e^{-{\bm\gamma}t}$,
we can write explicitly the expression of the average force 
$\langle F\rangle$ as
\begin{equation}
\begin{array}{lll}
\langle F\rangle & = &\langle F^*\rangle+V(N_p+\mu)=V(N_p+\mu)\left[1-e^{-{\bm\gamma}t}\left|_{11}\right.\right]+kVe^{-{\bm\gamma}t}\left|_{12}\right.\\  
&&\\
 & = & V(N_p+\mu)\left[1+\dfrac{\lambda_1e^{-\lambda_2t}-\lambda_2e^{-\lambda_1t}}{\lambda_1-\lambda_2}\right]
+kV\left[\dfrac{e^{-\lambda_1t}-e^{-\lambda_2t}}{(\lambda_1-\lambda_2)t}\right]. 
\end{array}
\end{equation} 
where $\lambda_{1,2}$ are the eigenvalues of ${\bm\gamma}$.
The only case in which we can have fluctuations in the average force is
obtained for $(\gamma^2-4\omega_0^2)<0$, that leads to the following
expression
\begin{equation}
\langle F\rangle = V(N_p+\mu)\left\{1+e^{-\frac{\gamma}{2}t}\left[\dfrac{\gamma}{2}\dfrac{\sin(\sqrt{4\omega_0^2-\gamma^2}t/2)}{\sqrt{4\omega_0^2-\gamma^2}/2}-\cos(\sqrt{4\omega_0^2-\gamma^2}t/2)\right]\right\}-kVe^{-\frac{\gamma}{2} t}\dfrac{\sin(\sqrt{4\omega_0^2-\gamma^2}t/2)}{\sqrt{4\omega_0^2-\gamma^2} t/2}.
\end{equation} 
In this case oscillations (serration) emerges, but they decays 
exponentially fast. On the other hand, performing the stationary limit,
one finds 
\begin{equation}\label{equ:stationary}
\lim_{t\rightarrow\infty}\langle F\rangle=V(N_p+\mu).
\end{equation}

Therefore, for any parameter values condition, a serration type
behavior is not observed in the single dislocation in a cloud of
mobile impurities model. 

%
%

\section{Dislocation pileup interacting with mobile impurities}

As the PLC effect is widely believed to be due to the dynamic interaction
of dislocations with diffusing solute atoms, a natural
formulation of the problem, in the framework of DDD approach, is to
consider Eq.~\ref{equ:laurson} for $N$ 
dislocations in a landscape of $N_p$ mobile impurities. 
To describe the dynamics of dislocations we use, as in
\cite{laurson2008}, an overdamped equation, so that the velocity of
a dislocation depends linearly on the resolved shear stress exerted
on it \cite{kanninen1969}.
Therefore, the equations of motion Eq.~\ref{equ:laurson} are
generalized as follows  
\begin{equation}\label{equ:plc}
\begin{array}{rll}
\mu\mbox{\Large{$\frac{d x_i}{d t}$}} & = &
G\begin{array}{l}
\\
\mbox{\Large{$\sum$}}_{j=1}^{N}\\
\mbox{\scriptsize{$(j\ne i)$}}
\end{array}
\dfrac{b_ib_j}{x_i-x_j}+b_i\sigma_
i^{l}+\mbox{\Large{$\sum$}}_{j=1}^{N_p}f_P(x_i-x_{s,j})\\
&&\\
\chi\mbox{\Large{$\frac{d x_{s,j}}{d t}$}} & = &
-\mbox{\Large{$\sum$}}_{i=1}^{N}f_P(x_i-x_{s,j})+\eta_j ,
\end{array}
\end{equation} 
where $G$ is the shear modulus, $b_i$ is the Burgers vector of the
dislocation $i$, $\mu$ and $\chi$ are the damping constant of
dislocations and impurities respectively.
The external force $F$ is now explicitly indicated as the local shear
stress $\sigma_i^{l}$ acting on each dislocation $i$, whose expression
is
\begin{equation}\label{equ:s_int}
\displaystyle \sigma_i^l=k\left[Vt-\int_{0}^{t} b_i\dfrac{d x_i(t')}{d t'}dt'\right]=k[Vt-b_i(x_i(t)-x_i(0))],
\end{equation} 
while for the pinning force $f_P(z)$, with $z=x_i-x_{s,j}$, and the noise term $\eta_j$ we have the
expressions 
\begin{equation}\label{equ:f-noise}
\begin{array}{rll}
f_P(z) & = & -f_0\dfrac{z}{\xi_P}\
\ \mbox{\large{$e^{-(z/\xi_P)^2}$}} ,\\
&&\\
\langle\eta_j(t)\rangle & = & 0 ,\\
&&\\
\langle\eta_j(t)\eta_j(t')\rangle & = & D\delta(t-t')\delta_{ij} .
\end{array}
\end{equation}  
To emulate the behavior of a material in the bulk, we consider that
$N$ point dislocations and $N_p$ impurities move along a line of size
$L$ where periodic boundary conditions are chosen. 
In order to correctly take into account the effect of periodic
boundary conditions, the interactions between dislocations are summed
over theirs images \cite{hirth1982}.

We are interested in the total average stress exerted on
dislocations, $\sigma=1/N\sum_{i=1}^{N}\sigma_i^l$ (i.e. the external
stress that is needed to apply on the material
to obtain a constant strain rate).
In view to study the behavior of the stress $\sigma$ in relation with
the PLC effect, that is regulated principally by the temperature and
strain rate as discussed in the introduction, we have imposed the
relations $b_i=b=1$, $G=\mu=\xi_P=1$, $f_0=0.01$ and $k=0.1$ that
fixes the time, space and force scales. 
The free parameters of the model are now $V, \chi, D$. 
In real materials impurities have already exerted aging effects over
dislocations before the experiments (i. e. before that an external
stress is imposed on the material).
To take into account this effect the system is leaved to evolve without
external stress ($k=0$) for a waiting time $t_w$.
The initial configuration of the system (at time $t=0$) consists of
random distribution of dislocations and pinning centers. 
We choose $t_w=10^6\cdot dt=10^4$, where the integration step is 
$dt=10^{-2}$.
This value of $t_w$ is sufficient for the system to stabilize its
elastic energy during the initial part of the dynamics (for $t<t_w$
and $k=0$) \cite{zaiser2006}.

\section{The conditions for serration}

Before integrating numerically Eqs.\ref{equ:plc}, \ref{equ:s_int},
\ref{equ:f-noise}, we can obtain a set of necessary
conditions for serration. 
First of all we can observe that if dislocations do not interact with
any pinning center ($f_P(z)=0$), the first equation of
Eq.\ref{equ:plc} do not posses normal modes of oscillation.
This can be found employing a linear perturbative approach, as made
in Ref.\cite{marcos2006} to study the discrete cosmological N-body
problem, or observing that the first equation of Eq.\ref{equ:plc}
describes the so called Coulomb gas for the variables $x_i-(V/b)t$ at
zero temperature \cite{dyson1962}.  

If we now consider that dislocations interact with pinning centers
($f_P(z)\ne0$), but with the last ones quenched (that means
$dx_{s,j}/dt=0$), from Eqs.\ref{equ:plc}, \ref{equ:s_int},
\ref{equ:f-noise} we obtain 
\begin{equation}\label{equ:quench}
\begin{array}{lll}
\begin{array}{rll}
\displaystyle \mu\dfrac{d x}{d t} & = &
\displaystyle b\sigma+\frac{1}{N}\sum_{i=1}^{N}\sum_{j=1}^{N_p}f_P(x_i-x_{s,j})\\
&&\\
x_{s,j} & = & c_j
\end{array}
\displaystyle \ \ \ \ \Longrightarrow \ \ \ \ \partial_t\sigma & = & \displaystyle kV -\dfrac{kb^2}{\mu}\sigma+\Gamma(t,\sigma^l_i,c_j)
\end{array}
\end{equation} 
where $c_j$ are constants and the function $\Gamma$ can be
obtained using the relation between $x_i$ and $\sigma^l_i$ in
Eq.~\ref{equ:s_int}.  
Performing the ensemble average and the time integral on the relation 
to the right in Eq.~\ref{equ:quench}, we obtain that 
$\langle\sigma\rangle=\mu V/b^2$ (so we do not have serration) if 
$\langle\Gamma\rangle=0$, that is if the constants $c_j$ do not 
correlate the variables $\sigma^l_i$ between them
(i.e. if the $c_j$ do not depend in a specific way on the position of
the variables $x_i(0)$).
To find this result we employed the relation $\sum_{i=1 (i\neq
  j)}^{N}\sum_{j=1}^{N}1/(x_i-x_j)=0$, introduced the variable
$x=1/N\sum_{i=1}^{N}x_i$ and considered the ensemble average respect
to the $\{c_j\}$ configurations.

Another necessary condition to have serration can be found observing
that in the range of parameters for which the impurities have a
vanishingly small probability to escape from dislocations (the
same case studied in Ref.\cite{laurson2008} for the single dislocation
problem) we can approximate the expression of the interacting force
$f(z)$ for small $z$ as $f_P(z)\simeq f_P(0)+z\partial_zf_P(z)|_{z=0}=-f_0z$. 
Employing this approximation, from Eqs.\ref{equ:plc}, \ref{equ:s_int},
\ref{equ:f-noise}, we find for $\sigma$ the following 2d
Ornstein-Uhlenbeck equation  
\begin{equation}\label{equ:S_total}
\partial_t^2\sigma=-\dfrac{1}{\mu}\left[kb^2+f_0N_p+\dfrac{f_0\mu N}{\chi}\right]\partial_t\sigma-\dfrac{f_0kb^2N}{\mu\chi}\sigma+\dfrac{f_0k}{\mu}\left[V(N_p+\dfrac{\mu N}{\chi})+\dfrac{kb^3N}{\chi}x(0)\right]-\dfrac{f_0kb}{\mu\chi}\sum_{i=1}^{N_p}\eta_i
\end{equation}
that do not displays serration as seen above.
To obtain the Eq.\ref{equ:S_total}, the relations 
$\sum_{i=1 (i\neq j)}^{N}\sum_{j=1}^{N}1/(x_i-x_j)=0$ and 
$\sum_{i=1 (i\neq j)}^{N}\sum_{j=1}^{N}\partial_t(x_i-x_j)/(x_i-x_j)^2=0$ has been employed.

\begin{figure}[h!]
\begin{center}
\includegraphics[clip=true,width=12cm]{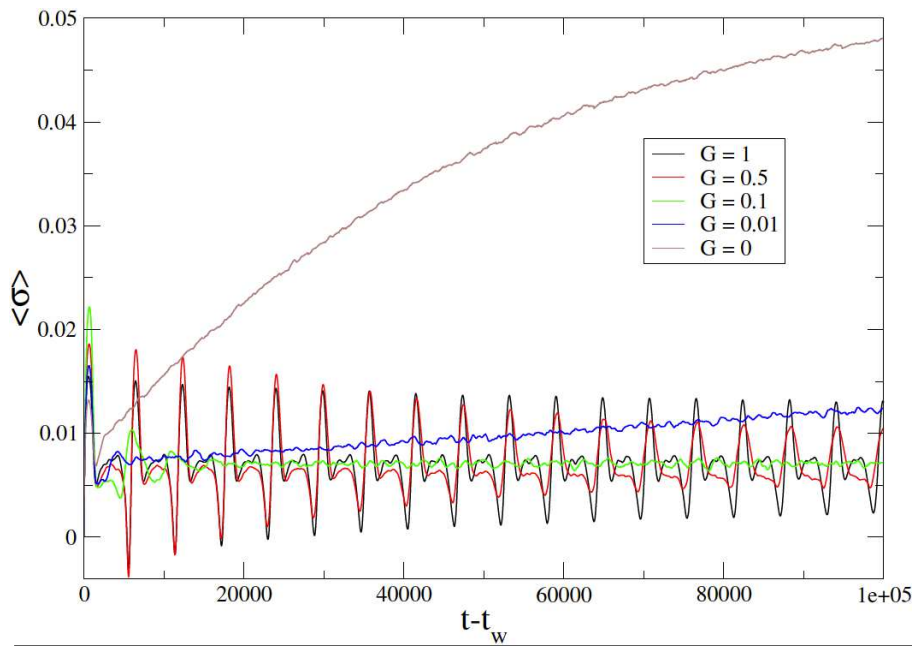}
\end{center}
\caption{Average stress $\langle\sigma\rangle$ (performed on 50
  samples) as a function of time obtained integrating 
numerically Eqs.\ref{equ:plc},\ref{equ:s_int}.  
The parameters of the system ($\mu/\chi=0.5$, $1/T=10^4$, and $V=0.003$) are chosen in 
the range in which serration is observed in order to investigate the role of the interaction 
force between dislocations changing the value of the shear modulus $G$.
From it we can see that serration disappear when the strength of the interaction force 
between dislocations decreases.}
\label{fig:Fig1} 
\end{figure}

\begin{figure}[h!]
\begin{center}
\includegraphics[clip=true,width=12cm]{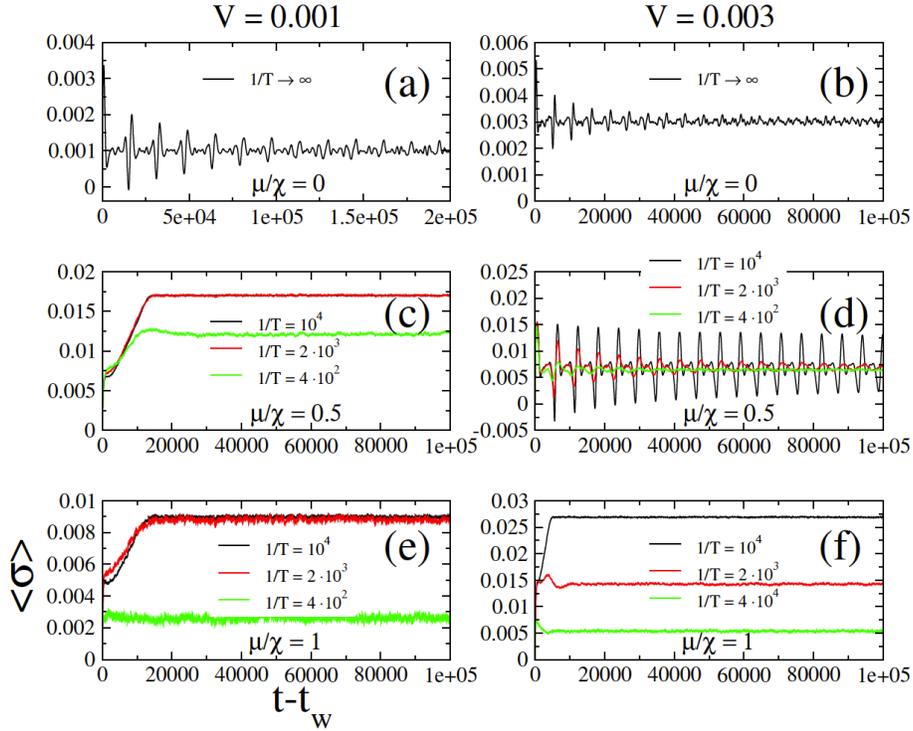}
\end{center}
\caption{Average stress $\langle\sigma\rangle$, obtained integrating numerically 
Eqs.~\ref{equ:plc},\ref{equ:s_int}, as a function of time $t-t_w$, for 
$t_w=10^6 \cdot dt=10^4$, driving velocity $V=0.001, 0.003$, mobility ratio 
$\mu/\chi=0, 0.5, 1$ and inverse temperature $1/T\rightarrow\infty$, 
$1/T=10^4, 2\cdot10^3, 4\cdot10^2$.}
\label{Fig2} 
\end{figure}

In the end, from Eqs.\ref{equ:plc}, \ref{equ:s_int}, \ref{equ:f-noise}
we have found four necessary conditions to have serration: {\it i)}
first of all, the system must be composed by more than one dislocation
(as found in Ref.\cite{laurson2008} in the low noise limit, and as
results in the general case from Eq.\ref{equ:stationary}). In other
words, serration in the stress response of the system, when present,
comes from a collective effect of many interacting dislocations. To
verify this, we studied the role of the interaction force between
dislocations analysing the average stress $\langle\sigma\rangle$ as a 
function of time (or strain) for different values of the interaction 
force itself (obtained changing the value of the shear modulus $G$), 
for parameter values in which serration is observed ($\mu/\chi=0.5$, 
$1/T=10^4$, see the definition of $T$ below, and $V=0.003$).
In Fig.\ref{fig:Fig1} we displayed the average stress
$\langle\sigma\rangle$ (performed on 50 samples) as a function of time
obtained integrating numerically
Eqs.\ref{equ:plc},\ref{equ:s_int}. From it we can see that serration
disappears when the strength of the interaction force between
dislocations decreases. In particular, for vanishing interaction force ($G=0$),
the stationary average stress is given by the expression: 
$\langle\sigma\rangle_s=V[(N_p/N)\chi+\mu]\simeq0.051$, as discussed in the
following analysis of Fig.\ref{Fig2}; 
{\it ii)} dislocations must interact with pinning centers; {\it iii)} pinning
centers must not be all quenched (see Eq.\ref{equ:quench}); {\it iv)} dislocations must not
be all pinned by impurities (see Eq.\ref{equ:S_total}).
These conditions are all in agreement with the DSA concept.

\begin{figure}[h!]
\begin{center}
\includegraphics[clip=true,width=8cm]{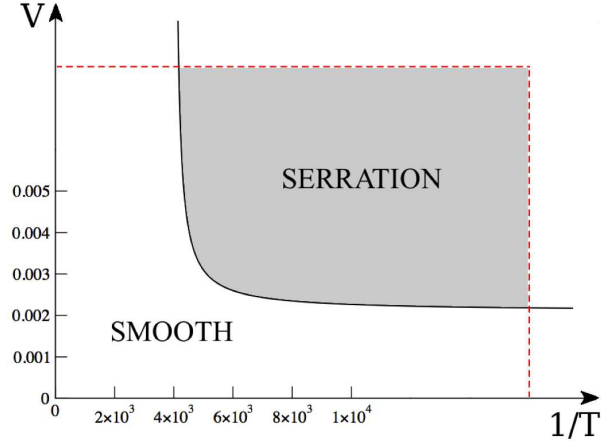}
\end{center}
\caption{Semi-quantitative phase diagram for the PLC effect in the 
parameter space ($1/T, V$) for $\mu/\chi=0.5$. In the range of
parameters inside the grey region, the PLC effect take place (i. e. the average stress $\langle\sigma\rangle$ displays serration).
The dashed red lines indicate where the model described by 
Eqs.~\ref{equ:plc},\ref{equ:s_int} start to fail in predicting the PLC effect.}
\label{Fig3} 
\end{figure}

In order to verify the previous four necessary conditions to have serration, 
and to investigate the behavior of the system in the entire free parameter 
space ($V,\chi,D$), we integrated numerically Eqs.\ref{equ:plc},\ref{equ:s_int}.
We considered $N=32$ dislocations with an average spacing $d=16$ and average 
pinning center spacing $d_p=L/N_p=2$. Because of $L=dN$, is $N_p=dN/d_p=8N$. 
Instead of using the set of parameters ($V,\chi,D$), we employed this other 
($V,\mu/\chi,1/T$) where $T$ is the temperature of the system defined in
the following. Although we choose, without less of generality, $\mu=1$ we 
prefer to keep explicitly the ratio $\mu/\chi$. 
From the other hand, the explicit introduction of the temperature
variable permits to compare the phase diagram of PLC effect (in
Fig.~\ref{Fig3}) with results from literature \cite{lebyodkin2000}.
The definition of temperature is not something obvious in DDD models.
Here we introduce the temperature $T$ of the system as obtained from the 
Einstein fluctuation-dissipation relation: $T=2D/\chi$.
One difficulty in deriving a general theory of plasticity is due to the 
presence of thermal as well as athermal dislocation activated processes 
\cite{ananthakrishna2007}.
For this reason is still lacking a clear definition of temperature in DDD 
models. The simple one we use here can be considered a good definition for 
high temperatures and low stresses in which cases diffusional deformation 
mechanisms become predominant \cite{zaiser2006}. In other regimes in which
this definition of $T$ is not a good approximation, we will discuss how it 
is related to the behavior of the system.

In Fig.~\ref{Fig2} the average stress $\langle\sigma\rangle$, 
obtained integrating numerically Eqs.~\ref{equ:plc},\ref{equ:s_int}, as a 
function of time $t-t_w$ is graphicated for driving velocity $V=0.001, 0.003$, 
mobility ratio $\mu/\chi=0, 0.5, 1$ and inverse temperature 
$1/T\rightarrow\infty$, $1/T=10^4, 2\cdot10^3, 4\cdot10^2$. 
The average $\langle\cdot\rangle$ is performed on 50 samples.
In the analysis of Fig.~\ref{Fig2} we distinguish two cases, 
the first one for $\mu/\chi=0$ and the second one for $\mu/\chi>0$.
The case $\mu/\chi=0$ corresponds to quenched pinning centers 
($dx_{s,j}/dt=0$, see Eq.~\ref{equ:quench}).
From Fig.~\ref{Fig2}a,b we see that for $\mu/\chi=0$, the 
average stress, after oscillations decreasing 
in time, reach the stationary value
$\langle\sigma\rangle_s=\lim_{t\rightarrow\infty}\langle\sigma\rangle=V\mu$.

In the case of $\mu/\chi>0$ we expect that at low temperatures and low 
driving velocity, or at low temperatures and high driving velocity, but 
high values of mobility ratio, dislocations are pinned by impurities.
This is confirmed by Fig.~\ref{Fig2}c,e,f from which
we can see that the stationary value of the average stress is 
given by the relation $\langle\sigma\rangle_s=V[(N_p/N)\chi+\mu]$, as can
be obtained from Eq.~\ref{equ:S_total}.
To understand the role of dislocation interaction when they are pinned
during the whole dynamics, we can consider the
Eq.~\ref{equ:stationary} obtained for the dynamics of a single
dislocation generalized to the case in which the damping constant of
each impurity is $\chi$.
Therefore the expression of the stationary external force becomes:
$\lim_{t\rightarrow\infty}\langle F\rangle=V(N_p\chi+\mu)$, in which  
$N_p$ is the number of impurities around the only present dislocation.
In the case of $N$ dislocations pinned by $N_p$ impurities during the whole 
dynamics, we have that on average each dislocation $i$ is pinned by $N_p/N$ 
impurities (considering that the initial spatial distribution of dislocations 
and pinning centers is a random flat ones). 
If we now suppose that in these conditions (obstacles that cannot
unpin from dislocations), dislocations do not fluctuate too much
around theirs equilibrium positions (that is a configuration of
equidistant dislocations), we can conclude that the average stationary
stress can be obtained from the formula for the single dislocation
pinned by $N_P/N$ obstacles case.
This means: $\langle\sigma\rangle_s=\langle F(N_p\rightarrow
N_p/N)\rangle_s$ that can be verified employing the previous
generalized version of Eq.~\ref{equ:stationary}.
When $T$ or $V$ increases, the number of impurities that pin dislocations
decreases, so $\langle\sigma\rangle_s$ decreases.

From Fig.~\ref{Fig2} we can see that stationary fluctuations 
in the stress (serration) emerge for the values parameter $V=0.003$, 
$\mu/\chi=0.5$ and $1/T=10^4$.  
In Fig.~\ref{Fig2} results for driving velocities higher than
$V=0.003$ and temperatures $T$ smaller than $10^{-4}$ are not reported
because in these regimes we reach the limit of our model that continue to
give serration in the stationary average stress, while in real systems we
should not have serration \cite{ananthakrishna2007, lebyodkin2000}.
Negative values of stress fluctuations at earliest times correspond to a 
sudden increases in the dislocations average position (see
Eq.~\ref{equ:s_int}) that happens when dislocations escape from many
pinning centers. Indeed, when it happens, we can observe this effect at
earliest times of the dynamics (for $t>t_w$), because the absence of
the external stress for $0<t<t_w$ permits to dislocations to
accumulate pinning centers. 

In order to summarise the results obtained from the present model and to 
compare them with experimental \cite{lebyodkin2000} and others theoretical 
approache \cite{ananthakrishna2007}, we depicted in Fig.~\ref{Fig3},
relying on data displayed in Fig.~\ref{Fig2} and others data not
displayed there, a phase diagram for the PLC effect in the parameter space 
($1/T, V$) for $\mu/\chi=0.5$.
In the range of parameters inside the grey region, the PLC effect takes place.
The dashed red lines indicate where the model described by 
Eqs.~\ref{equ:plc},\ref{equ:s_int} start to fail in predicting the PLC effect.
In particular, for high driving velocities new dislocation mechanisms, as
dislocation multiplication, climbing and others complex behaviors, must
be taken into account. Considering these mechanisms, serration must disappear 
for high driving velocities irrespective of the other parameter values.
While in the case of small temperatures, non thermal dislocation processes
become relevant in relation to the thermal ones, and the Einstein 
fluctuation-dissipation relation do not holds anymore. Also in this case, 
considering the presence of non thermal dislocation processes, serration must 
disappear for small temperatures irrespective of the other parameter values.

\vspace{1cm}

%
%

\section{The relevant time scales in serration}

\noindent The behavior of the system described by
Eqs.~\ref{equ:plc},\ref{equ:s_int},\ref{equ:f-noise} can be analyzed in 
terms of time scales.  
The fundamental time scales are: {\it i)} the capturing time $t_c$ that 
accounts for the average time needed for a dislocation to capture a
pinning center; 
{\it ii)} the aging time $t_a$ that accounts for the average time
needed for a dislocation to escape from a pinning center. 
\begin{table}[tp]
\caption{Time scales ratio: $t_a/t_c$. The computational error is 
  one over the last digit.}
\label{time_scales}\centering
\setlength{\extrarowheight}{1.1pt}
\begin{tabular}{cp{0.3cm}|c|lcp{0.3cm}|c|l}
\multicolumn{1}{l}{} & \multicolumn{1}{l}{} & \multicolumn{1}{c}{$V=0.001$} & & &\multicolumn{1}{c}{}& \multicolumn{1}{c}{$V=0.003$} & \multicolumn{1}{l}{} \\
\cline{3-3}
\cline{7-7}
\multicolumn{1}{l}{$\mu/\chi=0$} & {\hspace{0.11cm}\bm{$\{$}} & 0.60 & \footnotesize{$\leftarrow$ for all $1/T$} & \multicolumn{1}{l}{\hspace{0.5cm}$\mu/\chi=0$} & {\hspace{0.11cm}\bm{$\{$}} & 0.60 & \footnotesize{$\leftarrow$ for all $1/T$}\\
\cline{3-3}
\cline{7-7}
\multicolumn{1}{l}{} & \ldelim\{{3}{9mm} & 65.07 & \footnotesize{$\leftarrow 1/T=10^4$} & & \ldelim\{{3}{9mm}& {\bf 0.75} & \footnotesize{$\leftarrow 1/T=10^4$}\\
\multicolumn{1}{l}{$\mu/\chi=0.5$} & & 66.48 & \footnotesize{$\leftarrow 1/T=2\cdot 10^3$} & \multicolumn{1}{l}{\hspace{0.5cm}$\mu/\chi=0.5$}& & {\bf 0.75} & \footnotesize{$\leftarrow 1/T=2\cdot 10^3$}\\
\multicolumn{1}{l}{} & & 4.09 & \footnotesize{$\leftarrow 1/T=4\cdot 10^2$} & & & 0.64 & \footnotesize{$\leftarrow 1/T=4\cdot 10^2$}\\
\cline{3-3}
\cline{7-7}
\multicolumn{1}{l}{} & \ldelim\{{3}{9mm} & 78.73 & \footnotesize{$\leftarrow 1/T=10^4$} & & \ldelim\{{3}{9mm} & 146.97 & \footnotesize{$\leftarrow 1/T=10^4$}\\
\multicolumn{1}{l}{$\mu/\chi=1$} & & 37.93 & \footnotesize{$\leftarrow 1/T=2\cdot 10^3$} & \multicolumn{1}{l}{\hspace{0.5cm}$\mu/\chi=1$}& & 2.03 & \footnotesize{$\leftarrow 1/T=2\cdot 10^3$}\\
\multicolumn{1}{l}{} & & 1.13 & \footnotesize{$\leftarrow 1/T=4\cdot 10^2$} & & & 0.79 & \footnotesize{$\leftarrow 1/T=4\cdot 10^2$}\\
\cline{3-3}
\cline{7-7}
\end{tabular}
\end{table}
In order to compute these times we consider that a pinning center pins
a dislocation if the distance between them is smaller then 
$\xi_c=3\xi_P=3$
(otherwise the attraction force between them is considered negligible).  
In Tab.~\ref{time_scales} we report the time ratio
$t_a/t_c$ for the same values of driving velocity $V$, mobility
ratio $\mu/\chi$, and inverse temperature $1/T$ for which the average
stress $\langle\sigma\rangle$ has been computed and displayed in
Fig.~\ref{Fig2}. 
First of all, we can observe that for $V\rightarrow\infty$ or
$\mu/\chi\rightarrow 0$ or $1/T\rightarrow 0$, the time ratio become:
$t_a/t_c=(2\cdot\xi_c)/(d-2\cdot\xi_c)=0.6$ where $d$ is the dislocation 
average interdistance.   
Indeed, in these conditions we can consider the pinning centers to be
fixed respect to dislocations (or vice versa) during the dynamics and
the ratio $t_a/t_c$ become nothing more than the ratio between the
average (in time) length per dislocation and pinning
center over which dislocations are considered pinned (that is
$\alpha(t_{av})\cdot 2\cdot\xi_c$) and that over which
they are not (that is $\alpha(t_{av})\cdot(d-2\cdot\xi_c$))
where $\alpha(t_{av})$ is the same parameter for the two lengths and
depends only on the time $t_{av}$ over which the average is performed.
Looking at the values displayed in Tab.\ref{time_scales}, we can observe
that in general the time ratio $t_a/t_c$ decreases significantly as $V$ 
increases, or as $\mu/\chi$ or $1/T$ decreases, but only for 
$\mu/\chi=0.5$ and $V=0.003$ we have that $t_a/t_c$ remains small 
(bigger than, but near, the value $0.6$) for changing $1/T$.
In particular, for $\mu/\chi=0.5$, $V=0.003$ and $1/T=10^4, 2\cdot
10^3$, the two times $t_a$ and $t_c$ (which ratio is displayed in bold
on Tab.\ref{time_scales}) are of the same order of magnitude, but the
pinning centers, for these parameter values, are not fixed respect to
dislocations (or vice versa).  

In the end we can conclude that for values of parameters $V$,
$\mu/\chi$ and $1/T$ for which serration is observed, we have that the
two relevant times $t_a$ and $t_c$ are of the same order of magnitude
(remembering that the value $0.6$ corresponds to a special case in
our model).
This is in agreement with phenomenological models
\cite{ananthakrishna2007} and represents a link between microscopic
DDD model parameters (appearing in
Eqs.~\ref{equ:plc},\ref{equ:s_int},\ref{equ:f-noise}) from which
the quantities $t_a$ and $t_c$ can be computed, and macroscopic
quantities like $V$, $\mu/\chi$ and $T$.

%
%

\section{Spatio-temporal distribution of dislocations and impurities}

\begin{figure}[h!]
\begin{center}
\includegraphics[clip=true,width=12cm]{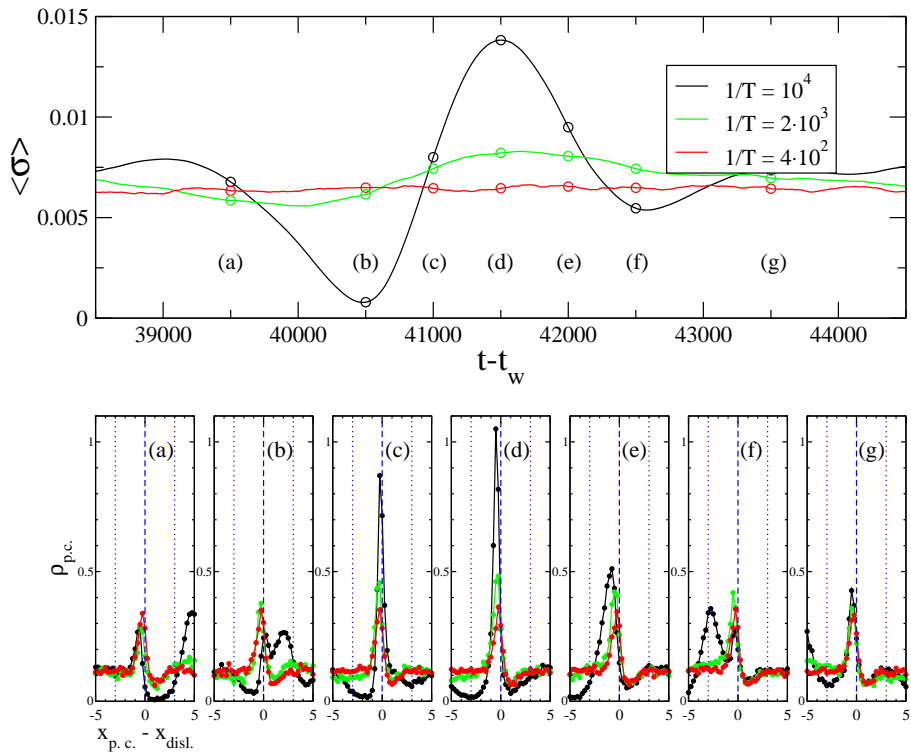}
\end{center}
\caption{Spatio-temporal distribution of pinning centers around dislocations.
The distributions are obtained averaging for a specific time the different 
distributions around every dislocation, and averaging over 50 different 
realizations of the dynamics. In the lower par of the figure are indicated
with $x_{p.c.}$ the pinning centers position and with $x_{disl.}$ the dislocation
position. The vertical dashed blue line indicate the dislocation position and
the two vertical dotted blue lines indicate the cutoff $\xi_c$ of the 
dislocation-pinning centers interaction. In the upper part of the figure,
the average stress for a specific time region is indicated. The corresponding 
complete curve is reported in Fig.~\ref{Fig2}.}
\label{Fig4} 
\end{figure}

Analysing spatio-temporal distributions of pinning centers and dislocations 
can help to better understand and unify previous considerations and results.
First of all, we computed the average pinning centers distribution $\rho_{p.c.}$ 
averaging each distribution of pinning centers around every dislocation, and 
averaging over 50 different realizations of the dynamics.
In the lower part of Fig.~\ref{Fig4} we display average 
pinning center distributions around dislocations for different times that 
correspond to different value of the average stress (which values are 
indicated in the upper stress vs time graphic of the same figure).
The more interesting case, reported in Fig.\ref{Fig4} 
and corresponding to Fig.\ref{Fig2}(d), is that for which changing 
temperature causes serration appearance or disappearance (that is for 
$\mu/\chi=0.5, V=0.003$).

The distributions $\rho_{p.c.}$ are unnormalized so that the average
number of pinning centers that pins dislocations are obtained by the integral: 
$\int_{-\xi_c}^{\xi_c}\rho_{p.c.}(x_{p.c.})dx_{p.c.}=\langle N_p\rangle$.
To understand what happens when serration in the stress appears, we reported
in Fig.~\ref{Fig4} the distributions in correspondence
of stress drop, bump and re-drop.
For temperatures for which the stress do not develops serration 
(for $1/T=4\cdot 10^2, 2\cdot 10^3$), the distribution shape of
pinning centers around dislocations do not change in time and displays
a peak more or less wide and narrow depending on the values of $V$,
$\mu/\chi$ and $1/T$.
For temperatures for which serration develops (for $1/T=10^4$), the 
distributions change in a way that when pinning centers reach dislocations from 
the right side (indeed $V=0.003>0$), Fig.~\ref{Fig4}(b),
and start to pin them, stress drops because  
dislocations are accelerated (see Eq.~\ref{equ:s_int}), then after a while 
dislocation velocities go down because they are pinned (when the corresponding 
distribution has developed a wide narrow peak), 
Fig.~\ref{Fig4}(c),(d). Finally, when dislocations 
are able to depin and then their velocities start to increase again, the 
stress re-drops, Fig.~\ref{Fig4}(e),(f). 
Above, analysing the stationary value of the stress (Fig.~\ref{Fig2}),
we argued that at low temperatures (in Fig.\ref{Fig2}(c),(e),(f)) 
dislocations are completely pinned, and verified that by means of the formula 
$\langle\sigma\rangle_s=V[(N_p/N)\chi+\mu]$ previously obtained from the equations 
of motion imposing the condition of complete pinning.
Integrating the distributions of pinning centers, $\rho_{p.c.}$, for
low temperatures, we can obtain the average number of pinning centers
around dislocations that is $\langle N_p\rangle=N_p/N=8$ which
confirms directly that in these cases all pinning centers are pinning dislocations.
Computing the standard deviation of the average distribution in
Fig.\ref{Fig4}, confirms that when serration
emerges, for $1/T=10^4$, dislocations depin in a choerent way.
Indeed, in this case the standard deviation is big near the peak of
the distribution, while is quite small away from it.
In the cases in which serration is not observed, for $1/T=10^4,4\cdot
10^2$, the standard deviation is almost constant in time and space.
Considering this result and that, in particular for $1/T=4\cdot 10^2$, the time 
ratio $t_a/t_c$ (see Tab.~\ref{time_scales}) is the same as for the case 
$1/T=10^4$, and that the average curve is constant in time, we can conclude 
that when serration disappear, this means that dislocations depin in a 
incoherent way (at different times). 

\begin{figure}[h!]
\begin{center}
\begin{minipage}{16.5cm}
\centering
\raisebox{-0.1\height}{\includegraphics[clip=true,width=8cm]{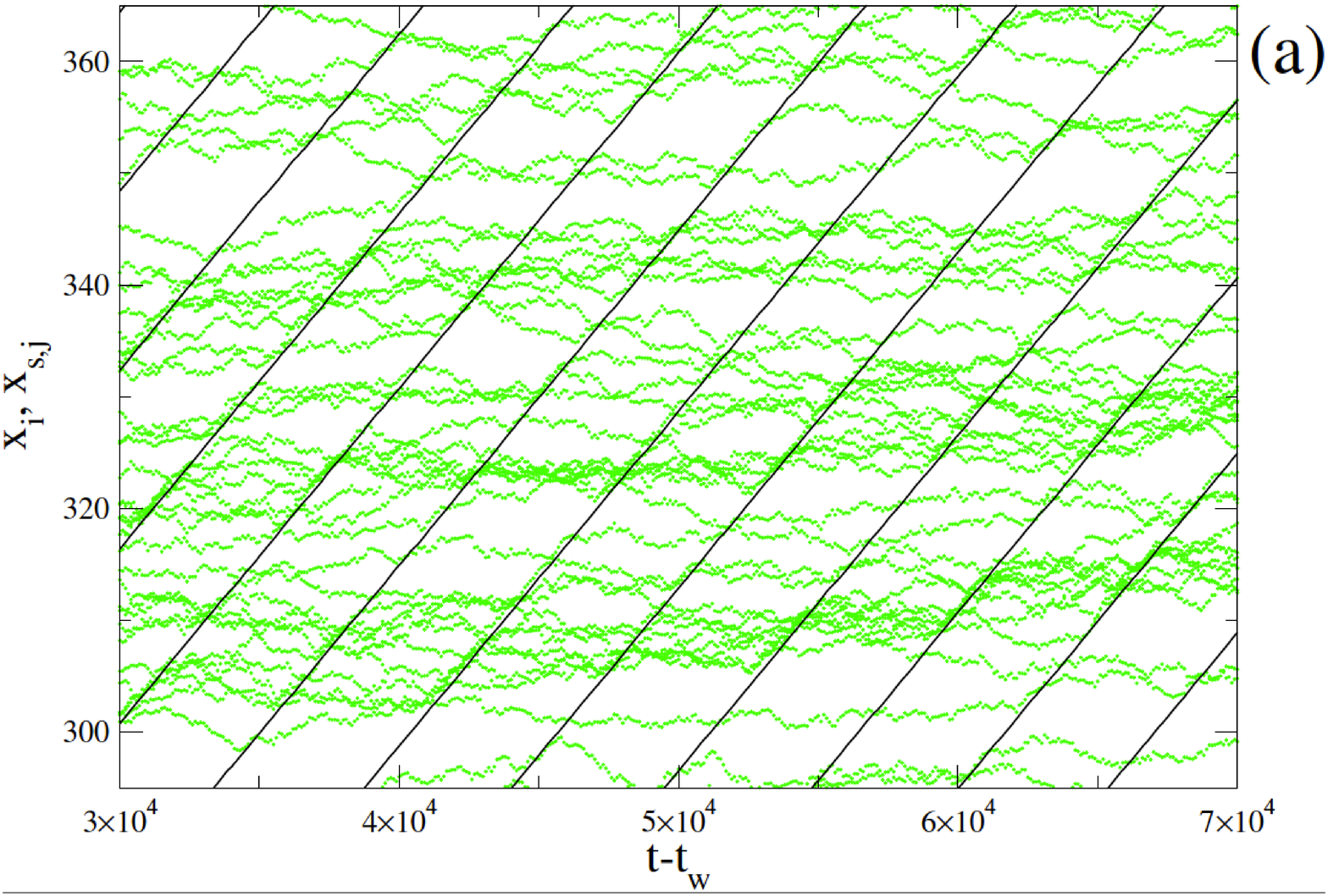}}
\hspace*{0.15cm}
\raisebox{-0.1\height}{\includegraphics[clip=true,width=8cm]{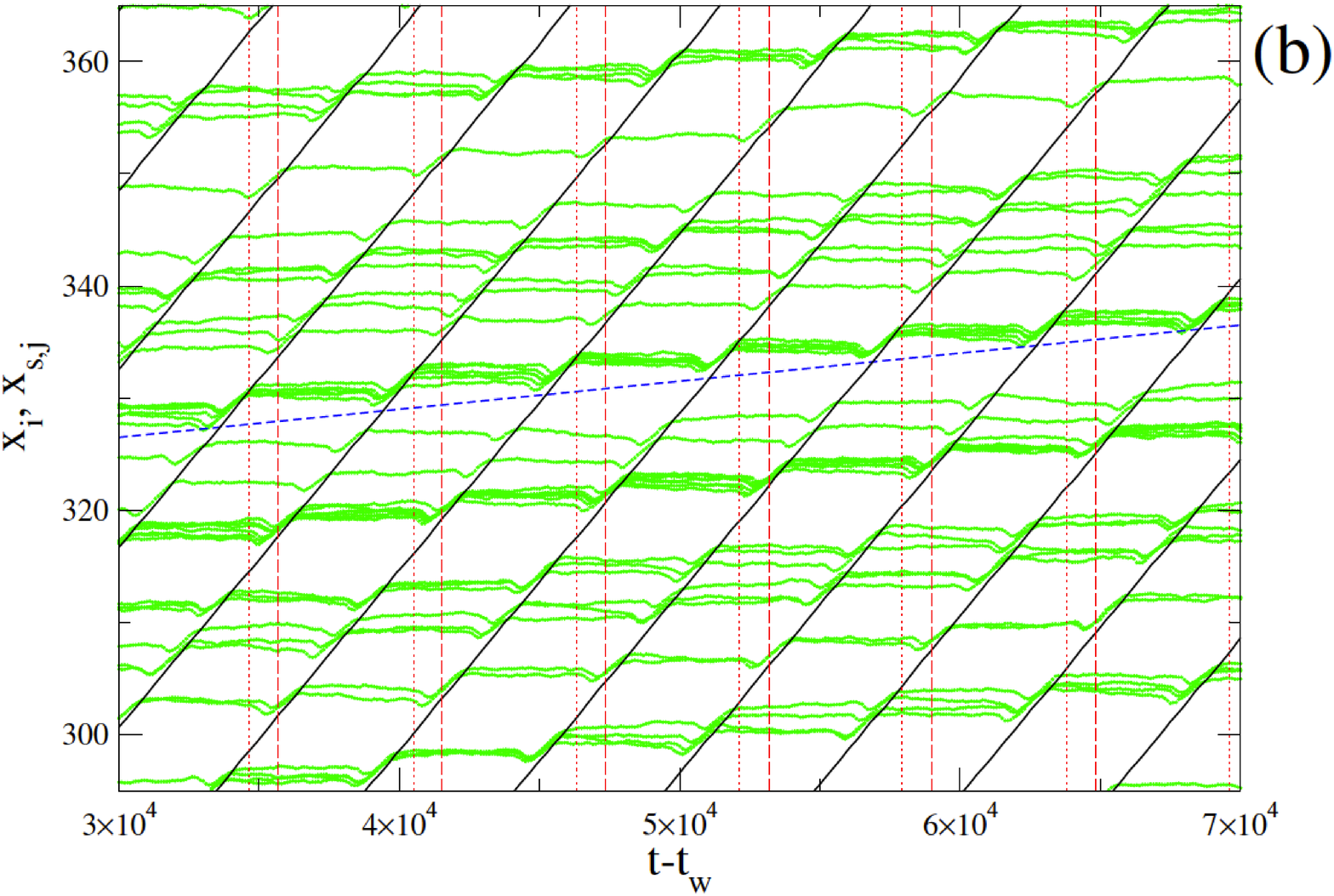}}
\end{minipage}
\end{center}
\caption{Representative portion of dislocations and pinning centers motion for
a single realization of the dynamics for parameter values: $V=0.003, 
\mu/\chi=0.5$ and $1/T=4\cdot 10^2$ (a), $1/T=10^4$ (b). 
The dark lines with a slope equal to V = 0.003 are dislocations, and the
green lines are pinning centers.
(b) The vertical dotted and dashed red lines correspond to times for which the 
average stress has a minimum and a maximum respectively. 
The dashed blu line $y\sim v_{drift}t$ indicates the drift of pinning centers that 
move with velocity $v_{drift}\simeq2.5\cdot10^{-4}$.}
\label{Fig5} 
\end{figure}
In Fig.~\ref{Fig5} we can see a representative portion of dislocations and
pinning centers motion in the case in which serration is not present
(a), and that in which it appears (b). 
The dark lines are dislocations. They have a slope equal to $V=0.003$, 
unless small deviations in correspondence of pinning centers (the green lines) 
pinning dislocations.
In Fig.~\ref{Fig5}b the vertical dotted and dashed red lines
correspond to times for which the average stress has a minimum and a
maximum respectively. 
The minimum in the average stress correspond to many dislocations that are 
pinned, while the successive maximum to many dislocations that depin.
These processes are also present at intermediate times, that cause the 
appearance of small oscillations in the average stress between a big peak and 
a big valley (see Fig.~\ref{fig:Fig1} and Fig.~\ref{Fig2}(d)).
The dashed blu line, $y\sim v_{drift}t$, indicates the drift of pinning centers 
that move with velocity $v_{drift}\simeq2.5\cdot10^{-4}$. This drift is caused by 
the interaction with dislocations.

\begin{figure}[h!]
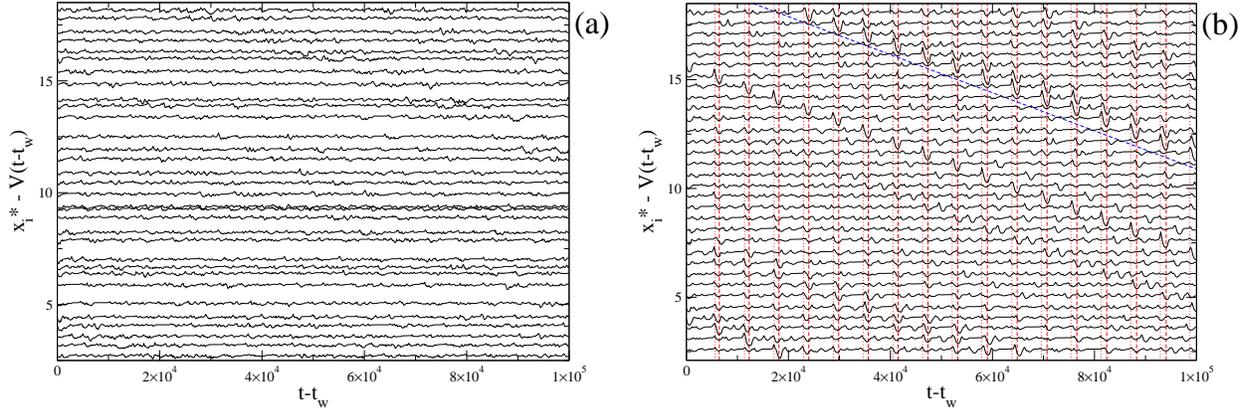

\begin{center}
\begin{minipage}{16.5cm}
\centering
\raisebox{-0.1\height}{\includegraphics[clip=true,width=8cm]{Fig6a.eps}}
\hspace*{0.15cm}
\raisebox{-0.1\height}{\includegraphics[clip=true,width=8cm]{Fig6b.eps}}
\end{minipage}
\end{center}
\caption{Dynamics of all dislocations for a single realization of the dynamics for 
parameter values: $V=0.003, \mu/\chi=0.5$ and $1/T=4\cdot 10^2$ (a), $1/T=10^4$ (b). 
To evidence fluctuations, dislocations position, at which the term
$Vt$ has been subtracted, are rescaled in a way that the average
inter-distance between them is 0.5 instead of $d=16$. 
(b) The dashed blue line $y\sim-0.5/[d/(V-v_{drift})]x=-0.875\cdot10^{-4}x$ indicate
clustering properties of pinning centers, as discussed in the text.}
\label{Fig6} 
\end{figure}
In Fig.~\ref{Fig6} we can see the entire dynamics of
all dislocations in a case in which there is not serration (a), and
another in which it  appears (b).
In particular, the term $Vt$ is subtracted to dislocation positions, so we have 
horizontal lines, while in Fig.~\ref{Fig5} we have lines with a slope equal to $V$. 
Moreover, the position of dislocations are rescaled in order to have that the average
inter-distance between them is 0.5 instead of $d=16$.
This last rescaling has been done to evidence the fluctuations of dislocations position.
Like in Fig.~\ref{Fig5}, the vertical dotted and dashed red lines correspond to times 
for which the average stress has a minimum and a maximum respectively.
From Fig.~\ref{Fig6}b we can see essentially two interesting things:
minimum and maximum values in the rescaled dislocations position correspond 
to maximum and minimum values in the average stress respectively; and the fluctuation
profile of the position of one dislocation (in correspondence of serration) can propagate 
with a velocity of $V-v_{drift}$.
This last point means that when pinning centers start to clustering, they
remain usually clustered during the entire dynamics.  
The dashed blue line $y\sim-0.5/[d/(V-v_{drift})]x=-0.875\cdot10^{-4}x$ in 
Fig.~\ref{Fig6}b indicate this behavior of pinning centers.

\begin{figure}[h!]
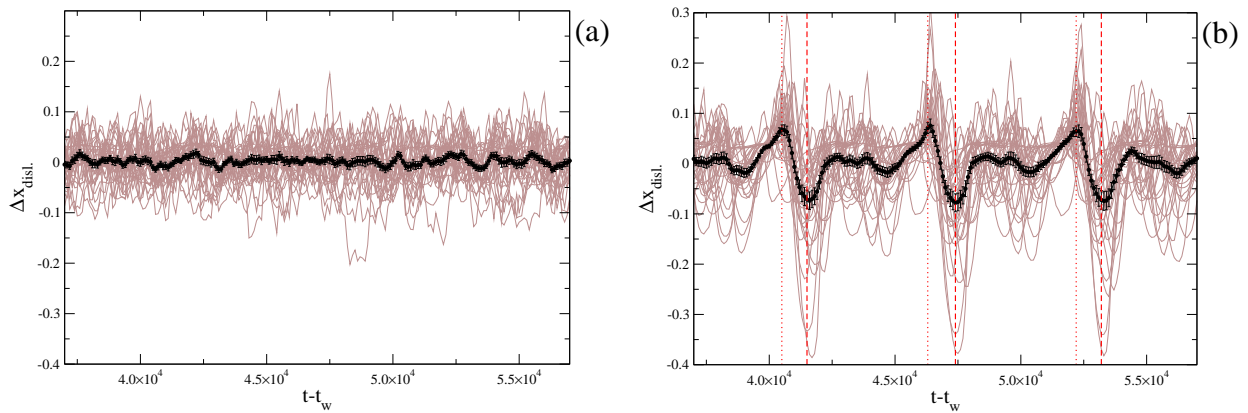

\begin{center}
\begin{minipage}{16.5cm}
\centering
\raisebox{-0.1\height}{\includegraphics[clip=true,width=8cm]{Fig7a.eps}}
\hspace*{0.15cm}
\raisebox{-0.1\height}{\includegraphics[clip=true,width=8cm]{Fig7b.eps}}
\end{minipage}
\end{center}
\caption{In brown we display the superposition of the position 
fluctuations of all dislocations, $\Delta x_i=(x_i-Vt)-<(x_i-Vt)>_t$ where the 
positions $x_i$ are reported in Fig.~\ref{Fig6}, and the time average 
is performed on the entire dynamics.
In dark we can see the average of all $\Delta x_i$ with the relative standard 
deviation. (b) The vertical dotted and dashed lines correspond to times for which 
the average stress has a minimum and a maximum respectively.}
\label{Fig7} 
\end{figure}

In Fig.~\ref{Fig7} we display the superposition of fluctuations 
in the position of all dislocations, $\Delta x_i$, in a case in which
there is no serration (a) and another in which it appears (b).
These fluctuations has been obtained as: $\Delta
x_i=(x_i-Vt)-\langle(x_i-Vt)\rangle_t$, where the time average is performed on the 
entire dynamics. 
We also display the average of all $\Delta x_i$ with the relative
standard deviation, and the vertical dotted and dashed lines
correspond to times for which the average  
stress has a minimum and a maximum respectively.
We can see another time that the minimum and maximum values of the average 
fluctuation position correspond to the maximum and minimum in the average stress 
respectively. 

%
%

\section{Discussion}

We investigated the dynamics of a dislocation assembly interacting
with mobile impurities studying the case of a one-dimensional dislocation
pileup. In order to connect this model to the PLC effect, we
studied the stress response of the system under an external constant
strain rate.  The free parameters of the system have been reduced to the driving
velocity $V$, that controls the imposed constant strain rate, the mobility
ratio $\mu/\chi$ between dislocations and mobile impurities, and the temperature $T$.
To this end, we have employed an effective definition of temperature that should be
valid except for low temperatures and high stresses. 
Analysing the average stress of the system $\langle\sigma\rangle$ in
the parameter space ($V, \mu/\chi, 1/T$), we found a region characterized
by stationary fluctuations in the stress (serration)  (see
Fig.~\ref{Fig3}).

The interpretation for the onset of serration in the present model
agrees with the general concept of DSA but takes explicitly into account
the role of dislocation mutual interactions.
The emergence of serration corresponds to the situation in which 
impurities diffuse at a rate that allows them to pin dislocations (i.e.
the capturing time $t_c$ is not too high). At the same time
dislocations should be able to escape from pinning centers after an
aging time $t_a$ which implies that $t_a$ is not too high, otherwise when
dislocations unpin, they would move at large speed until they are arrested
again and pinning centers will not be able to reach them, but also 
not too small, because in this case pinning centers will not be
able to pin dislocations. Serration correspond to the case in which the
 two characteristic time scales, $t_a$ and $t_c$, are of the 
same order of magnitude (see
Tab.~\ref{time_scales}).   

From another point of view, the origin of serration, in a specific
range of parameter values ($V, \mu/\chi, T$), is due to the localization 
of impurities in  a limited number of clouds under the action of dislocation
induced stresses and to the possibility for dislocations to escape from their
pinning clouds without randomizing excessively the spatial distribution
of impurities. The spatial localization of pinning centers 
is only possible due to the coherent action of several interacting dislocations.
If we randomize the interaction between dislocations, for instance
by  choosing  Burgers vectors $b_i$ for each dislocation $i$ from a random
distribution, serration disappears.    
Spatial randomization of pinning centers results in an incoherent 
contribution to the total stress fluctuations of each dislocation and 
in the impossibility to form clouds of pinning centers and therefore in the
disappearance of serration.  Only when the contribution of each dislocation to the total
stress fluctuations are coherent, serration is observed.

In order to extend the validity of the definition of temperature and
generalize the present model, the following
effects could be introduced: {\it i)} a viscosity therm for
dislocations; {\it ii)} pinning centers with different mobilities.
The last modification could give different type of serration like in
experiments in which three types of bands can be identified (type
A,B,C) \cite{ananthakrishna2007}.

\section*{Acknowledgments}
SZ is supported by the European Research Council through the Advanced Grant SIZEFFECTS.

\end{document}